\documentclass[proof]{pasj00}
\Received{$\langle$reception date$\rangle$}
\Accepted{$\langle$acception date$\rangle$}
\Published{$\langle$publication date$\rangle$}
\SetRunningHead{Interaction beween the Outflow and the Dense Gas}{Shimajiri et al.}

\usepackage{times}

\begin{document}

\title{Millimeter- and Submillimeter-Wave Observations\\ 
             of the OMC-2/3 region. IV\\
Interaction between the Outflow and the Dense Gas \\
in the Cluster Forming Region of OMC-2 FIR 6}

\author{YOSHITO SHIMAJIRI$^{1,4}$, SATOKO TAKAHASHI$^{2}$, SHIGEHISA TAKAKUWA$^{2}$, MASAO SAITO$^{3}$, and RYOHEI KAWABE$^{4,5}$ 
\thanks{Last update: May. 18, 2009}}
\affil{$^{1}$ Department of Astronomy, School of Science, University of Tokyo, Bunkyo, Tokyo 113-0033, Japan}
\affil{$^{2}$ Academia Sinica Institute of Astronomy and Astrophysics, P.O. Box 23-141, Taipei 106, Taiwan}
\affil{$^{3}$ ALMA Project Office, National Astronomical Observatory of Japan, Osawa 2-21-1, Mitaka, Tokyo 181-8588, Japan}
\affil{$^{4}$ Nobeyama Radio Observatory, Minamimaki, Minamisaku, Nagano 384-1805, Japan}
\affil{$^{5}$ National Astronomical Observatory of Japan, Osawa 2-21-1, Mitaka, Tokyo 181-8588, Japan}

\email{yoshito.shimajiri@nao.ac.jp}

\maketitle

\begin{abstract}
We have conducted millimeter interferometric observations of the Orion Molecular
Cloud-2 (OMC-2) FIR 6 region at an angular resolution of $\sim$ 4$\arcsec$ - 7$\arcsec$ with the 
Nobeyama Millimeter Array (NMA). In the 3.3 mm continuum emission we detected
dusty core counterparts of the previously identified FIR sources (FIR 6a, 6b, 6c, and 6d), 
and moreover, resolved FIR 6a into three dusty cores.
The size and mass of these cores are estimated to be 1100-5900 AU and 0.19-5.5 M$_{\odot}$,
respectively. We found that in the $^{12}$CO ($J$=1--0) emission
FIR 6b, 6c, and 6d eject the molecular outflow
and that the FIR 6c outflow also exhibits at least two collimated jet-like components
in the SiO ($J$=2--1) emission. At the tip of one of the SiO components there appears abrupt increase of the SiO line
width ($\sim$ 15 km s$^{-1}$), where the three resolved cores in FIR 6a seem to delineate the tip.
These results imply the presence of the interaction and the bowshock front
between the FIR 6c molecular outflow and FIR 6a.
\begin{bfseries}
If the interaction occurred after the formation of the FIR 6a cores
the influence of the FIR 6c outflow on the FIR 6a cores is minimal,  
since the total gravitational force in the FIR 6a cores (1.0 - 7.7 $\times$ 10$^{-4}$ M$_{\odot}$ km s$^{^1}$ yr$^{-1}$)
is much larger than the outflow momentum flux (2.4 $\times$ 10$^{-5}$ M$_{\odot}$ km s$^{^1}$ yr$^{-1}$).
On the other hand, it is also possible that
the interaction caused the gravitational instability in FIR 6a, and triggered the fragmentation into three cores, 
since the separation among these cores ($\sim$ 2.0 $\times$ 10$^{3}$AU) is on the same order of the Jeans length 
($\sim$ 5.0 - 8.4 $\times$ 10$^{3}$AU). In either case, FIR 6a cores, with a mass of 0.18 - 1.6 M$_{\odot}$ and a density of 
0.2 - 5.8  $\times$ 10$^{7}$ cm$^{-3}$, might be potential formation sites of the next generation of cluster members.
\end{bfseries}

\end{abstract}

\section{INTRODUCTION}

Most stars ($\geq$ 90 \%) form in the mode of cluster formation 
\citep{Lada03}.
However, as compared to the isolated star formation 
\citep{Sai99,Sai01,tak04,tak07b,Jes07}, the mechanism and the physical
process of the cluster formation are still less clear.
Previous observational and theoretical studies suggest that some 
external effects, such
as supernovae \citep{Kobayashi08} and HII regions \citep{Koening08}, 
are required
to trigger cluster formation.
\begin{bfseries}
Another candidate of such trigger is a 
molecular outflow,
as proposed both observationally \citep{San01,Yoko03} and
theoretically \citep{Naka07}.
The slight increase of the external pressure added by
the outflow makes the dense gas gravitationally unstable although it is still
close to the equilibrium state, and proceeds the
further gravitational fragmentation of the dense gas \citep{Elmegreen98},
which initiates the formation of stellar clusters \citep{Whitworth94}.
\end{bfseries}
The observational evidence of triggered star formation, however, have yet been mainly based on morphologies and the 
detailed discussions
of the physical processes are still less comprehensive.

The Orion Molecular Cloud-2/3 (OMC-2/3) region is one of the nearest 
cluster-forming
regions, including intermediate-mass protostars, that has been observed in the infrared, submillimeter, and 
millimeter bands. 
\citep{Gen89,Cas95,Chini97,Joho99,Yu00,Lada03,Tsuji03,Will03,Allen07}.
We have initiated survey observations of this region at millimeter and 
submillimeter wavelengths
with Nobeyama Millimeter Array (NMA) and Atacama Submillimeter 
Telescope Experiment (ASTE)
(see overview by \cite{Takaha08a}). In this survey, \citet{Takaha08} 
identified totally fourteen
molecular outflows in the $^{12}$CO ($J$=3--2) emission, including 
seven new detections.
Comparison between the 850 $\mu$m dust continuum emission 
\citep{Joho99} and our
$^{12}$CO ($J$=3--2) outflow results
provides two interesting candidates of the interaction between the 
dusty dense
gas and the outflow, namely, FIR-3/4 and FIR 6, both of which are
cluster-forming regions. On the other hand, in the MMS 7 region where 
there
is a single intermediate-mass protostar no observational evidence for 
the
outflow-dense gas interaction is found, and
the star-formation in MMS 7 is more likely to be in the isolated mode 
\citep{Takaha06}.

These findings have stimulated us to conduct
detailed studies of the interaction between molecular outflows
and dense gas as one of the possible mechanisms of cluster formation.
Our recent observations of the FIR 3/4 region \citep{Shima08} revealed 
that
the molecular outflow driven by FIR 3 interacts with the dense gas 
associated with FIR 4.
Furthermore, the 3.3 mm dust-continuum observations with the NMA
at a high spatial resolution ($\sim$ 3 $\arcsec$) have revealed that 
FIR 4 consists of
eleven dusty cores. 
From the comparison between the core separation and the Jeans length,
the fragmentation and the interacting time scale, and between the 
virial and LTE masses
of the dusty cores, we propose that the interaction triggered the 
fragmentation into
these dusty cores, and the next generation 
\begin{bfseries} of 
\end{bfseries}
cluster formation in FIR 4.

In order to investigate whether the scenario of
the outflow-triggered cluster formation is ubiquitous or a specific 
case only applicable to
FIR 4, it is necessary to search for more samples.
In this paper, we report detailed millimeter interferometric 
observations
of the other target, the FIR 6 region in OMC-2 \citep{Chini97}.
In the FIR 6 region, there are five MIR sources and one 3.6 cm VLA 
source \citep{rei99,nie03}.
In this region, we found two molecular outflows driven by FIR 6b and 6c 
\citep{Takaha08a,Takaha08},
and the outflow driven by FIR 6c appears to interact with the dense 
molecular gas at
FIR 6a seen in the 850 ${\mu}$m dust continuum emission \citep{Joho99}.
Since this situation of the FIR 6 region is similar to that of the FIR 
4 region,
the FIR 6 region is a potential site for the outflow-triggered cluster 
formation.

\section{OBSERVATIONS AND DATA REDUCTION}

We carried out millimeter interferometric observations of the FIR 6 region in the 
$^{12}$CO ($J$=1--0; 115.271 GHz), SiO ($v$=0, $J$=2--1; 86.847 GHz) lines and in the 3.3 mm continuum emission with the Nobeyama 
Millimeter Array (NMA), which consists of six 10 m antennas, during a period from 2006 January to 2008 February. 
The phase reference centers of the $^{12}$CO ($J$=1--0) observations are ($\alpha_{J2000}$, $\delta_{J2000}$) = (5$^{h}$ 35$^{m}$ 
23${\fs}$28, -5$^{\circ}$ 12$^{\arcmin}$ 3${\farcs}$19) and (5$^{h}$ 35$^{m}$ 23${\fs}$46, -5$^{\circ}$ 13$^{\arcmin}$ 15${\farcs}$20), while those of the SiO ($v$=0,$J$=2--1) and 3.3 mm continuum observations are ($\alpha_{J2000}$, $\delta_{J2000}$) = (5$^{h}$ 35$^{m}$ 23${\fs}$38, -5$^{\circ}$ 12$^{\arcmin}$ 19${\farcs}$19) and (5$^{h}$ 35$^{m}$ 20${\fs}$66, -5$^{\circ}$ 13$^{\arcmin}$ 15${\farcs}$00).
The molecular line data were obtained with the FX correlator, which was configured with 1024 channels per baseline and a bandwidth of 32 MHz. 
For the $^{12}$CO ($J$=1--0) and SiO ($v$=0, $J$=2--1) data we made 5 and 10-channel binning to increase the signal-to-noise ratio ($\equiv$ S/N) of the high-velocity line-wing emission, providing the velocity resolution in the $^{12}$CO and SiO observations of 0.406 km s$^{-1}$ and 1.08 km s$^{-1}$, respectively. 
 We obtained the continuum data at both the lower (87.090 $\pm$ 0.512 GHz) and upper (98.418 $\pm$ 0.512 GHz) sidebands with the digital spectral correlator, Ultra Wide Band Correlator (UWBC, \cite{oku00}). 
To obtain a higher S/N in the continuum map, the data of both sidebands were co-added (effective observing frequency = 92 GHz, 
corresponding to the wavelength of 3.3 mm). 

 Using the AIPS package developed at NRAO, we adopted the natural weighting for the imaging of the molecular emissions.
For the continuum imaging we adopted both the uniform and natural UV weighting.
Table \ref{line_obs} and Table \ref{cont_obs} summarize the parameters for the line and continuum observations, respectively. 
Our observations of the $^{12}$CO ($J$=1--0) and SiO ($v$=0, $J$=2--1) lines were insensitive to structures more extended than 42$\farcs$9 (0.09 pc) and 50$\farcs$0 (0.1 pc) 
at the 10 \% level \citep{Wilner94}, since the minimum projected baseline length of the $^{12}$CO ($J$=1--0) and 
SiO ($v$=0, $J$=2--1) observations was 3.9 and 3.3 k$\lambda$, respectively. 
The overall uncertainty in the flux calibration was estimated to be $\sim$ 15 \%. 
After the calibration, only the data taken under good weather conditions were adopted in the imaging.

\section{RESULTS}

\subsection{3.3 mm Dust Continuum Emission}

\begin{figure}
   \begin{center}
      \FigureFile(160mm,100mm){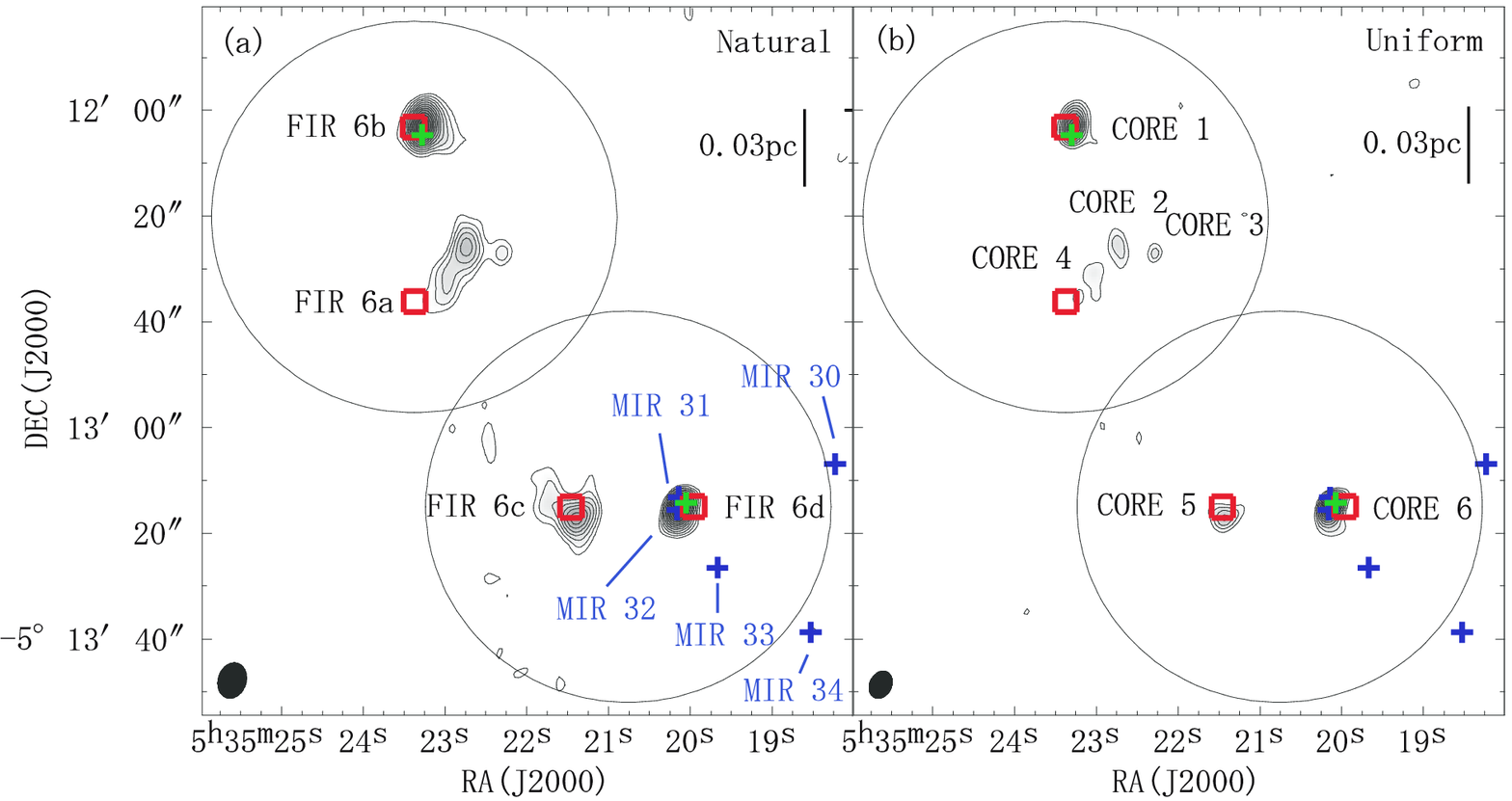}
   \end{center}
   \caption{3.3 mm dust continuum maps in the FIR 6 region observed with the NMA. 
The left panel and right shows the naturally-weighted map and uniformly-weighted map. 
Red squares show the position of the 1.3 mm dust continuum sources, FIR 6a, b, c, and d \citep{Chini97}. 
Blue crosses show the position of MIR 30,31,32,33, and 34 \citep{nie03}. 
Green crosses show the position of 24 $\mu$m sources (ID 36 and 38; \cite{Takaha08}). 
Ellipses at the bottom-left corner of each panel show the beam size.  
Open circles denote the field of view of the NMA observations. 
Contour levels of panel (a) and (b) start at $\pm$ 3 $\sigma$ 
noise levels with an interval of 1 $\sigma$. 
The rms noise level (1 $\sigma$) in panel (a) and (b) are 1.0 and 1.2 mJy beam$^{-1}$, respectively.
}\label{dust}
\end{figure}

Figure \ref{dust}a and \ref{dust}b show naturally- and 
uniformly-weighted 3.3 mm
dust continuum images in the FIR 6 region, respectively. To make these 
images,
we first CLEANed the South-West and the North-East field individually,
and then we corrected them for the primary beam response and combined 
them
with the Miriad task ``linmos''.
The rms noise levels before the primary beam correction are similar at 
both field, and we
adopted these noise levels as references of the contour levels.
 From these images, we identified six 3.3 mm continuum condensations 
with the following criteria;
(1) the peak intensity of condensations should be higher than the 3 
$\sigma$ noise level,
(2) the ``valley" among different condensations should be deeper than 1 
$\sigma$, and
(3) the structure is consistent between the naturally- and 
uniformly-weighted images,
and the size should be larger than the relevant beam size.
Hereafter we call these condensations CORE 1-6, as labeled in Figure 
\ref{dust}.

CORE 1, 5, and 6 are associated with FIR 6b, c, and d, which are previously
identified in the 1300 $\mu$m observations \citep{Chini97}. 
CORE 1 (FIR 6b) is also associated with a 24 $\mu$m source detected with Spitzer MIPS (ID 36; \cite{Takaha08}). 
CORE 5 is not associated with any MIR source, but shows a cavity-like structure pointing toward FIR 6a in the naturally-weighted 3.3 mm continuum image, 
whose direction is consistent with the direction of the outflow driven by FIR 6c (see Figure \ref{CO32}a and c). 
CORE 6 is also associated with a 24 $\mu$m source detected with Spitzer MIPS (ID 38; \cite{Takaha08}), as well as 
5, 10, and 20 $\mu$m sources (MIR 30 and 31; \cite{nie03}). 
This core has an elongated structure toward the SE-NW direction, which coincides with the direction of the molecular outflow driven by FIR 6d traced by the $^{12}$CO (1--0) emission (see Figure \ref{CO32}d). 
Toward FIR 6a, an elongated dusty structure is seen in the naturally-weighted 3.3 mm continuum image. 
In the uniformly-weighted image, we have resolved FIR 6a into three cores, CORE 2, 3, and 4. 
These cores do not harbor any MIR sources.  
Hereafter we call this aggregation FIR 6a clump. 

The deconvolved size of each dusty core was estimated by the 2-dimensional Gaussian fitting to the uniformly-weighted image, 
providing the typical deconvolved size of $\sim$ 1100 - 5900 AU. 
\begin{bfseries}
The total gas + dust mass
of the cores ($\equiv$ M$_{dust}$) was derived from the total 3.3 mm continuum flux (F$_{\nu}$)
in the uniformly-weighted image, on the assumption that all the 3.3-mm continuum emission arises from dusts,
the emission is optically thin, and that the gas-to-dust mass ratio is 100,\end{bfseries}
using the formula;
\begin{equation} 
M_{dust}=\frac{F_{\nu}d^2}{\kappa _{\nu}B_{\nu}(T_{d})}, \label{dustmass}
\end{equation}
where we adopt a value of the mass opacity, 
$\kappa _{\nu }=0.1 \left(\frac{250\mu m}{\lambda _{92 GHz}} \right) ^\beta$ cm$^{2}$ g$^{-1}$ \citep{Hildebrand83},
and
\begin{bfseries} distance $d$ = 400 pc \citep{Menten07,Sandstrom07}. \end{bfseries}
For the dust temperature we adopt the temperature range of T$_{d}$= 20 - 52.7 K.
The lower boundary, T$_{d}$ = 20 K, is adopted from the typical dust temperature of the OMC-3 cores \citep{Chini97} and 
the higher boundary, T$_{d}$ = 52.7 K, is derived from the peak intensity of the $^{12}$CO ($J$=3--2)
spectrum toward FIR 6 (see Figure \ref{spec}).

\begin{bfseries}
We adopt $\beta$ = 2 for the warm dusty cores in the FIR 6 region,
since FIR 6 is located in the same molecular filament of OMC-2 as FIR 1 and FIR 2, where the $\beta$
value is estimated to be 2 by the spectral 
energy distribution fitting \citep{Chini97}. 
The masses of CORE 1, 2, 3, 4,  5, and 6 are estimated to be 1.9 - 5.5, 0.56 - 1.6, 0.18 - 0.51, 0.52 - 1.5, 1.6 - 4.5, and 1.4 - 4.0 M$_{\odot}$, respectively. 
The mean gas density ($\equiv$$n$) in these dusty cores was derived by assuming a spherically-symmetric shape as follows;

\begin{equation}
n=\frac{M_{dust}}{\frac{4}{3}\pi\biggl(\sqrt{\frac{D_{maj}}{2}\times \frac{D_{min}}{2}}\biggl)^{3}}, \label{density}
\end{equation}

where D$_{maj}$, and D$_{min}$ are the deconvolved size along the major and minor axes. 
The typical gas density is $\sim$ 10$^{7}$ cm$^{-3}$. Table \ref{core} summarizes these physical properties of the identified cores.  

\end{bfseries}

\subsection{$^{12}$CO (1--0) Emission}
\begin{figure}
   \begin{center}
      \FigureFile(160mm,100mm){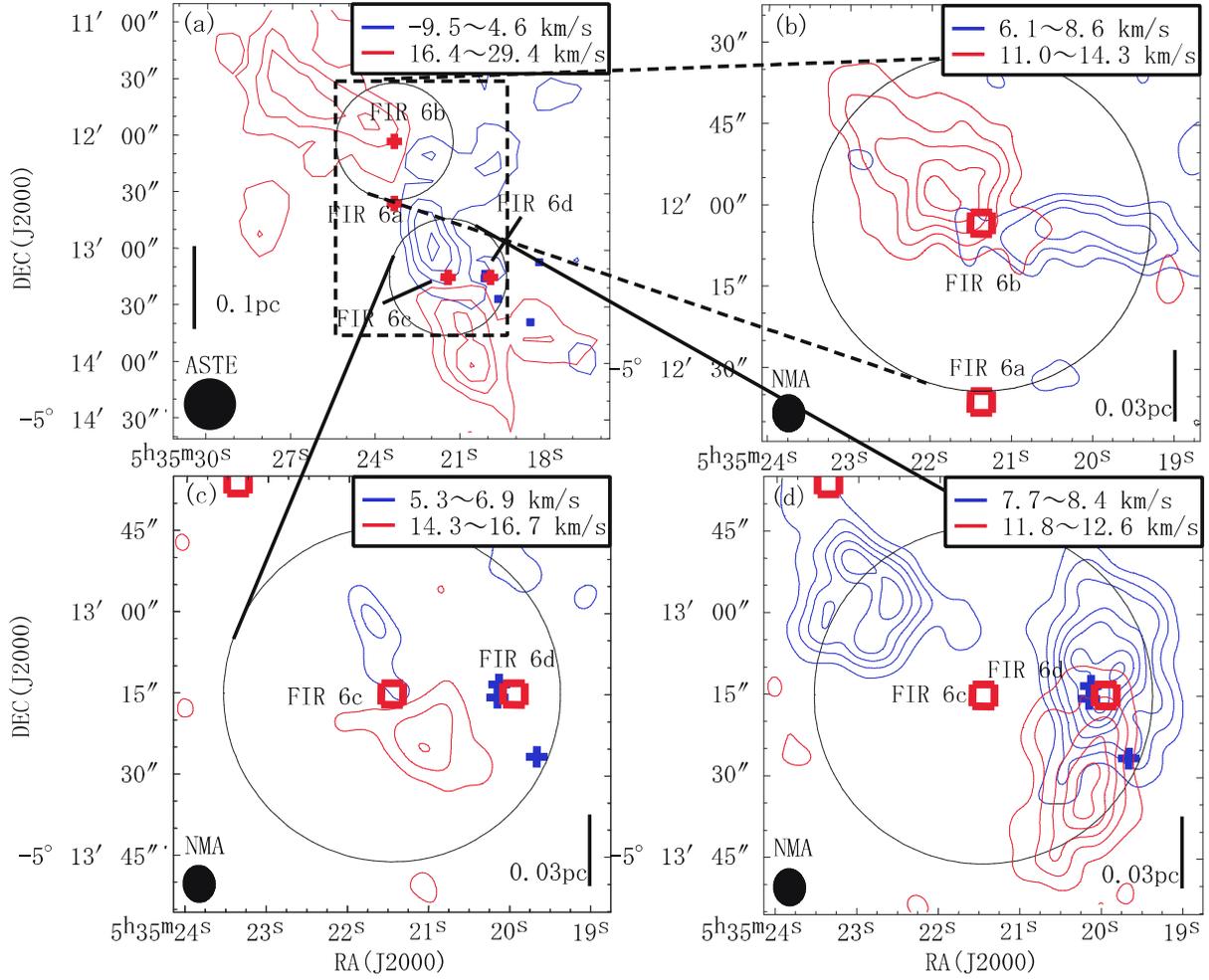}
   \end{center}
 \footnotesize
   \caption{Distribution of the high-velocity blueshifted (blue contour) and
redshifted (red contour) $^{12}$CO ($J$=3--2) (the upper left panel) and 
$^{12}$CO ($J$=1--0) emission (the other panels) in the FIR 6 region. 
Panel (a) shows the distribution of the blue (-9.5 km s$^{-1}$- 4.6 km s$^{-1}$) and 
red (16.4 km s$^{-1}$- 29.4 km s$^{-1}$) lobes in the $^{12}$CO (3--2) line observed with ASTE \citep{Takaha08}, 
while panel (b), (c), and (d) show the distribution of blue (6.1-8.6, 5.3-6.9, and 7.7-8.4 km s$^{-1}$) and 
red (11.0-14.3, 14.3-16.7, and 11.8-12.6 km s$^{-1}$) lobes in the $^{12}$CO (1--0) line observed with the NMA, respectively. 
Symbols in the figure are same as in Figure \ref{dust}.  
Contour levels of panel (a) start at $\pm$ 5 $\sigma$ levels with an interval of 5 $\sigma$, while 
contour levels of panel (b), (c), and (d) start at $\pm$ 3 $\sigma$ levels with an interval of 3 $\sigma$. 
The rms noise levels (1 $\sigma$) in the panel (a), (b), (c), and (d) are 1.2 $\times$ 10$^{3}$ K, 
0.11, 0.14, and 0.19 Jy beam$^{-1}$, respectively.
}\label{CO32}
\end{figure}

\begin{figure}
   \begin{center}
      \FigureFile(160mm,100mm){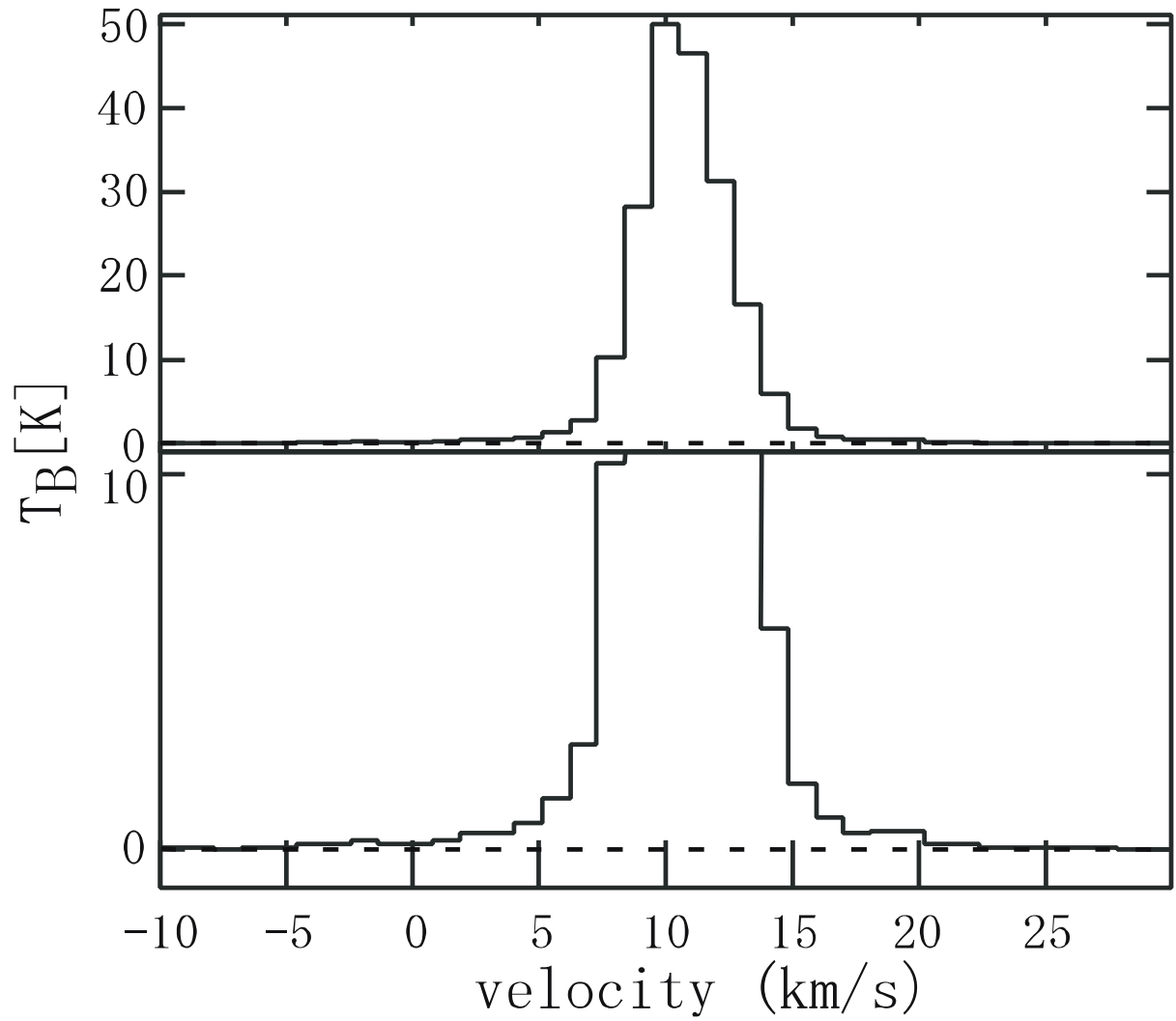}
   \end{center}
 \footnotesize
   \caption{$^{12}$CO ($J$=3--2; 345.705 GHz) spectrum in the FIR 6 region averaged over an area of 2.8$\arcmin \times 1.5\arcmin$ (see Figure \ref{CO32}) taken with ASTE \citep{Takaha08}. The top panel shows the entire spectrum and the bottom panel shows the 
enlarged view of the line-wing.
The rms noise level (1$\sigma$) is 
0.5 K in T$_B$ at a velocity resolution of 1.1 km s$^{-1}$. 
}\label{spec}
\end{figure}

Figure \ref{CO32}a shows distributions of the high velocity blueshifted (-9.5 $\sim$ 4.6 km s$^{-1}$ ) and 
redshifted (16.4 $\sim$ 29.4 km s$^{-1}$) $^{12}$CO ($J$=3--2) emission in the FIR 6 region. 
The $^{12}$CO ($J$=3--2) data were taken by \citet{Takaha08}. 
The total velocity range of the $^{12}$CO ($J$=3--2) emission is much wider than that of the H$^{13}$CO$^{+}$ ($J$=1--0) emission (9.8 $\sim$ 11.9 km s$^{-1}$; \cite{Ike07}), suggesting that the high velocity $^{12}$CO ($J$=3--2) emission traces the molecular outflows.  
Molecular outflows ejected from FIR 6b and 6c were identified by \citet{Takaha08},
although these molecular outflows were not detected in the previous single-dish $^{12}$CO ($J$=1--0) observations \citep{Aso00}. 
These molecular outflows were also detected in the NMA $^{12}$CO ($J$=1--0) observations as shown in Figure \ref{CO32}b and \ref{CO32}c, respectively. 
The NMA observations show a clearer bipolar morphology of the FIR 6b outflow. 
In particular, a narrow jet-like feature is seen in the blueshifted South-Western lobe. 
The intensity of the NMA $^{12}$CO ($J$=1--0) outflow driven by FIR 6c, on the other hand, is fainter than that of the ASTE $^{12}$CO ($J$=3--2) outflow, 
suggesting that the outflow driven by FIR 6c has extended components ($\geq $ 42$\farcs$9 $\sim$ 0.09 pc). 

Moreover, we have found a new outflow candidate in the NMA $^{12}$CO (1--0) observations, whose driving source is FIR 6d as shown in Figure \ref{CO32}(d). 
Since the velocity range of this outflow component detected with the NMA 
 is within that of the ambient cloud component observed with the single-dish ASTE telescope (8.9 km s$^{-1}$ to 13.2 km s$^{-1}$), this outflow component is hidden in the ambient cloud component and undetectable with ASTE. 
The distribution of the blueshifted and redshifted $^{12}$CO ($J$=1--0) emission around FIR 6d has a 
clear bipolar morphology, and the velocity range of the $^{12}$CO ($J$=1--0) emission (7.7 $\sim$ 12.6 km s$^{-1}$) is 
larger than that of the H$^{13}$CO$^{+}$ emission (9.8 $\sim$ 11.9 km s$^{-1}$, \cite{Ike07}). Hence, these bipolar components in the 
$^{12}$CO ($J$=1--0) emission associated with FIR 6d probably trace the molecular outflow.

Figure \ref{spec} shows the $^{12}$CO ($J$=3--2) line profile averaged within the region delineated by the dashed rectangle in Figure \ref{CO32}a. 
The high-velocity line wings, arising from the ensemble of these outflows, are evident in this profile.
From the peak temperature of this $^{12}$CO ($J$=3--2) spectrum, we determined that the upper limit of the gas kinetic temperature in the FIR 6 region is 52.7 K.

\subsection{SiO Emission}
\begin{figure}
   \begin{center}
      \FigureFile(160mm,100mm){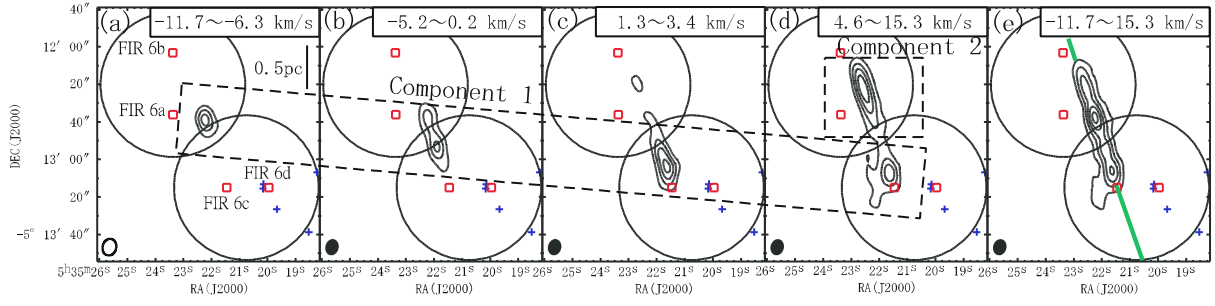}
   \end{center}
   \caption{Spatial and velocity distribution of the SiO ($v$=0, $J$=2--1) emission in the FIR 6 region observed with the NMA. 
Panel a, b, c, d, and e show the integrated intensity maps of the SiO emission for five different velocity ranges of 
-11.7 $\sim$ -6.3 km s$^{-1}$, -5.2 $\sim$ 0.2 km s$^{-1}$, 1.3 $\sim$ 3.4 km s$^{-1}$, 4.6 $\sim$ 15.3 km s$^{-1}$, 
-11.7 $\sim$ 15.3 km s$^{-1}$, and total, respectively. 
Symbols in the figure are same in Figure  \ref{dust}. 
The green line shows the cut line of the Position-Velocity diagram.
Contour levels start at $\pm$ 10 $\sigma$ with an interval of 10 $\sigma$. 
The rms noise level (1 $\sigma$) is 0.126, 0.131, 0.052, 0.151, and 0.225 Jy beam$^{-1}$ km s$^{-1}$ for panel a-e, respectivly.
}\label{SIO}
\end{figure}

Figure \ref{SIO}a-d, and e show integrated intensity maps of the SiO ($v$=0, $J$=2--1) emission at four different velocity ranges of -11.7 $\sim$ -6.3 km s$^{-1}$, -5.2 $\sim$ 0.2 km s$^{-1}$, 1.3 $\sim$ 3.4 km s$^{-1}$, 4.6 $\sim$ 15.3 km s$^{-1}$, and the 
total velocity range, respectively. 
The method of the multifield imaging is same as that for Figure \ref{dust}
In the total integrated intensity map of Figure \ref{SIO}e, a well collimated, linearly extended structure with a position angle of 20$^{\circ}$ is 
seen in the SiO emission. The South-Western tip of this SiO feature coincides with the position of FIR 6c. 
In the low velocity range shown in Figure \ref{SIO}d two SiO components can be identified; 
one is the compact SiO component located close to FIR 6c (Component 1), 
and the other extending along the North-East to South-West direction located close to FIR 6a and FIR 6b (Component 2). 
In the higher blueshifted velocity range, Component 1 appears to shift its peak position away from FIR 6c progressively, and 
the highest velocity peak locates close to FIR 6a (see Figure \ref{SIO}a). 
On the other hand, there is no such velocity gradient seen in Component 2. 

We will discuss the origin of these SiO components in section 4.1.

\section{DISCUSSION}

\subsection{Interaction between the Molecular Outflow Driven by FIR 6c and FIR 6a Clump}

\begin{figure}
   \begin{center}
      \FigureFile(160mm,100mm){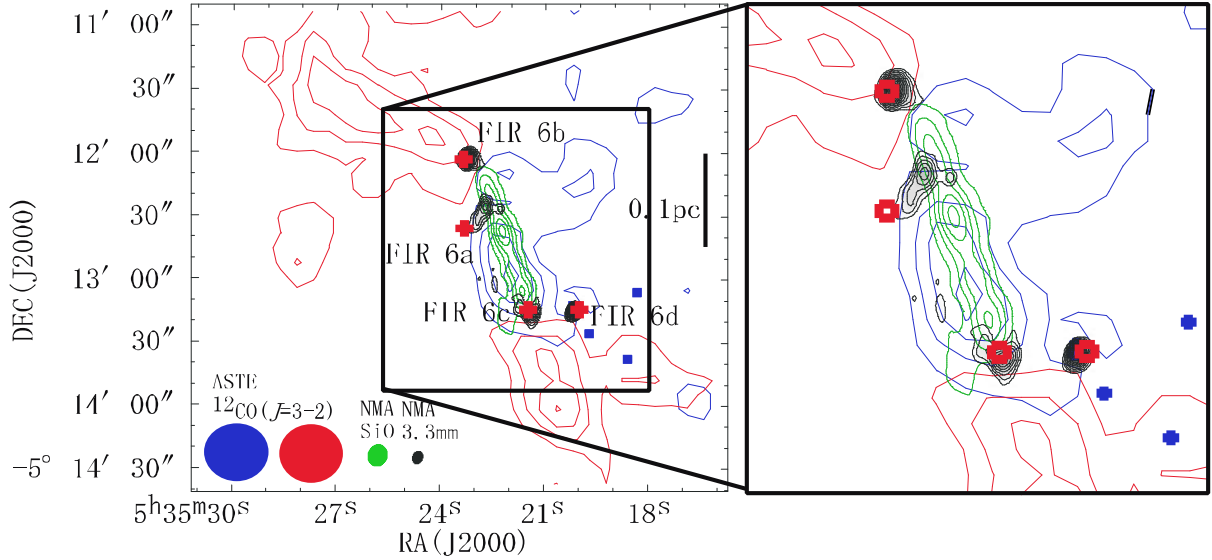}
   \end{center}
   \caption{Comparison of the spatial distribution of the $^{12}$CO ($J$=3--2), SiO ($v$=0, $J$=2--1), and the 3.3 mm dust continuum emissions in the FIR 6 region. Red and blue contours show the distribution of the redshifted (16.4 $\sim$ 29.4 km s$^{-1}$) and blueshifted (-9.5 $\sim$ 4.6 km s$^{-1}$) lobe in the $^{12}$CO ($J$=3--2) emission. Green contours show the distribution of the total integrated intensity map of the SiO emission. 
Black contours show the distribution of the 3.3 mm dust continuum emission. 
Contour levels of the $^{12}$CO ($J$=3--2) emission, SiO ($v$=0, $J$=2--1) emission and the 3.3 mm continuum emission are 
the same as in Figure \ref{CO32}a, \ref{SIO}e, and Figure \ref{dust}a, respectively.
Symbols in the figure are same as in Figure \ref{dust}. 
}\label{all}
\end{figure}

\begin{figure}
   \begin{center}
      \FigureFile(160mm,100mm){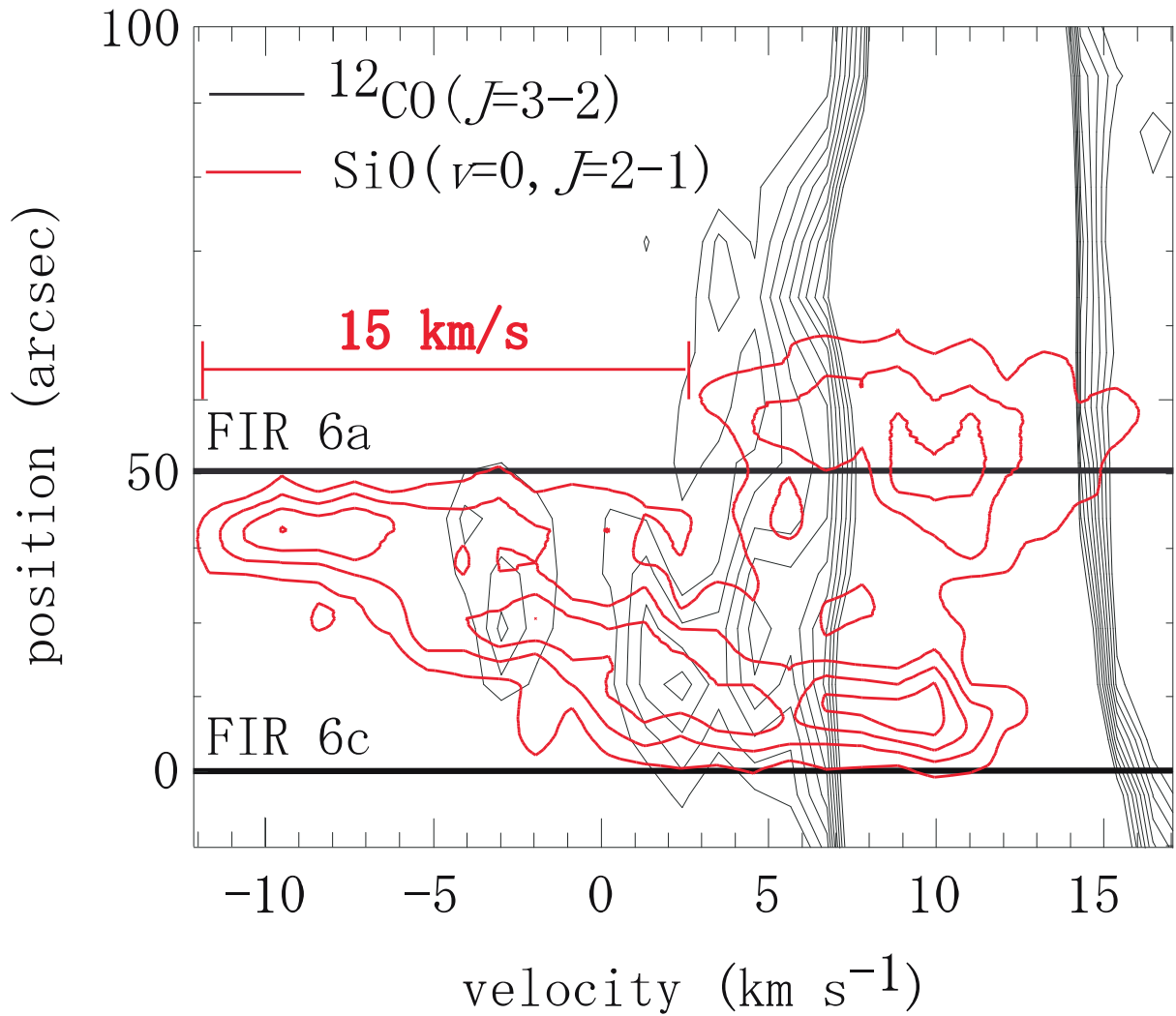}
   \end{center}
   \caption{Position-Velocity diagram of the $^{12}$CO ($J$=3-2; black contour) and the SiO emission (red contour) 
cut along the axis of the molecular outflow driven by FIR 6c (P.A. = 20$^{\circ}$). 
The vertical axis shows the positional offset with respect to the position of FIR 6c. 
Horizontal lines show the position of FIR 6a clump and FIR 6c, respectively. 
Contour levels start at $\pm$ 5 $\sigma$ with an interval of 2.5 $\sigma$. 
The rms noise level (1 $\sigma$) of the SiO and $^{12}$CO emission is 0.062 Jy beam$^{-1}$ and 0.46 K, respectively.
}\label{PV}
\end{figure}

In Figure \ref{all}, we compare the spatial distribution of the high-velocity $^{12}$CO ($J$=3--2) emission, SiO ($v$=0, $J$=2--1) emission, and the 3.3 mm dust continuum emission. The linearly extended feature in the SiO emission coincides well with the blueshifted lobe of the molecular outflow driven by FIR 6c. Since the SiO emission often traces shocked molecular gas caused by the interaction between the primary jet and the ambient molecular gas \citep{Avery96,Bach91,Hirano06}, this SiO emission probably traces the well collimated outflow driven by FIR 6c. 
Furthermore, 
CORE 2, 3, and 4 in FIR 6a clump surround 
the tip of Component 1 in the SiO emission, 
while the tip of Component 2 locates even more downstream beyond FIR 6a clump. 
In Figure \ref{PV}, we show Position - Velocity diagrams of the $^{12}$CO ($J$=3--2) and SiO ($v$=0, $J$=2--1) lines along the 
axis of the FIR 6c outflow. 
It is obvious that there are two distinct outflow components in the $^{12}$CO and SiO emission; 
the velocity component with the higher blueshifted velocity corresponds to Component 1, while the other lower velocity component located away from FIR 6c corresponds to Component 2. 
The SiO emission of Component 1 shows significant increase of the line width at the position of FIR 6a clump. 
This increase of the SiO line width, as well as the shell-like structure in FIR 6a clump at the interface
with Component 1, implies the presence of the interaction between
the molecular outflow driven by FIR 6c and FIR 6a clump \citep{tak03,Shima08}. 
We suggest that the two distinct SiO components were ejected 
toward slightly different 3-dimensional directions, as seen in other protostellar jets \citep{Vel98,Bach01}, and that the direction of Component 1 matches the direction to FIR 6a clump from FIR 6c. 
Therefore, the jet component 1 interacts with FIR 6a clump while Component 2 keeps moving freely. 
On the assumption of the same jet propagation velocity, the lower line of sight velocity of Component 2 implies that 
the jet axis of Component 2 is 
\begin{bfseries}
closer
\end{bfseries}
 to the plane of the sky than that of Component 1, which is consistent with the larger distance of Component 2 from FIR 6c.

\begin{bfseries}

Such an interaction associated with the outflow alters the physical condition of 
 the surrounding medium through the C-type shock \citep{Bergin98,Gusdorf08}.
\citet{Gusdorf08} have reported that a postshock density is 10 - 40 times more than a preshock density 
from their MHD simulation of the interaction. 
Since the present gas density of FIR 6a clump is estimated to be $\sim$ 1.0 - 2.9 $\times$ 10$^{7}$ cm$^{-3}$, 
the preshock density of FIR 6a clump is $\sim$ 2.6 - 7.3 $\times$ 10$^{5}$ cm$^{-3}$ 
on the assumption that a postshock density is 40 times more than a preshock density.

The redshifted counterpart of the SiO emission, on the other hand, was not detected. 
One of the possible reasons for the non-detection of the SiO emission is that there 
is not sufficient dense-gas material at the south-west of FIR 6c, and hence the reservoir 
of the SiO production or amount of dust grains \citep{Bach01} are not ample enough. 
In fact, the 1.3 mm dust continuum map in the FIR 6 region by \citet{Chini97} 
exhibits that there is asymmetric distribution of dusts around FIR 6c, 
and that at the south-west of FIR 6c the dust continuum emission becomes fainter 
while toward the north-east of FIR 6c the dust continuum emission becomes stronger 
and shows a emission ridge.

\subsection {Effect of the Interaction on the FIR 6a Cores}
In section 4.1., we demonstrate that the outflow driven by FIR 6c interacts with FIR 6a clump. 
Three cores in FIR 6a clump locate around the tip of the FIR 6c outflow, and it is possible
that the interaction affects the physical evolution of these cores. In the following section,
we discuss the possible effects of the interaction on these cores.

First, we compare the time scale of the interaction to the time scale of fragmentation of
FIR 6a clump to produce the three cores. On the assumption that the interaction time scale 
$\tau_{interaction}$ is similar to the dynamical time-scale of the FIR 6c outflow, i.e. $\tau_{interaction}$ $\sim$ 
$\tau_{d}$, $\tau_{interaction}$ is estimated to be 1.2-1.9 $\times$ 10$^4$ yr \citep{Takaha08}. The time scale of the 
fragmentation into the cores can be estimated on the assumption that the fragmentation time scale 
 is the sound crossing time; $\tau_{fragmentation} = {\frac{\Delta l}{C_{eff}}}$
$\sim$ 1.5 $\times$ 10$^{4}$ yr. These two times scales are indistinguishable from each other
and hence it is not straightforward to tell whether the interaction takes place before or after
the production of the three cores. Then we will discuss the following two cases;
\begin{itemize}
\item{CASE 1: The FIR 6a cores were formed before the interaction occurs.} 
\item{CASE 2: The FIR 6a cores were formed after the interaction occurs.}
\end{itemize}

In CASE 1, the FIR 6a cores could have been formed via spontaneous fragmentation or turbument fragmentation
before the interaction \citep{Inutsuka97,Li04}. In this case, the outflow driven by FIR 6c may
externally affect the subsequent evolution of these cores.
\citet{Maury09} have compared the total force required to balance the gravity of the dense cores
with the outflow momentum flux in the NGC 2264-C region, in order to assess whether
the outflows influence the evolution of the neighboring cores. We follow the same
argument for our case of the FIR 6 region to investigate whether the blueshifted component of the
FIR 6c outflow influences the physical evolution of the FIR 6a cores.
The total force needed to balance gravity $F_{grav}$ is expressed as 
$F_{grav}(R) = G \times M(R)^{2} / 2R^{2}$ (eq.10; \cite{Maury09}). 
Since the average radius of the FIR 6a cores and the mass are estimated to be 
$R$ = 0.0075 pc and $M(R)$ = 1.3 -3.6 M$_{\odot}$, respectively (see section 3.1. ),
$F_{grav}$ is calculated to be 1.0 - 7.7 $\times$ 10$^{-4}$ M$_{\odot}$km s$^{-1}$ yr$^{-1}$.
On the other hand, the outflow momentum flux of the blueshifted component of the FIR 6c outflow $F_{flow}$ is
2.4 $\times$ 10$^{-5}$ M$_{\odot}$km s$^{-1}$ yr$^{-1}$ \citep{Takaha08}, and hence
the total force is  $\sim$ 10 times larger than the outflow momentum flux ($F_{grav} >> F_{flow}$ ). 
Therefore, it is unlikely that the interaction with the FIR 6c outflow influences
the subsequent physical evolution of the FIR 6a cores.
\citet{Maury09} have also reported that the total momentum flux of eleven outflows in NGC 2264-C is 
insufficient to prevent the cores from collapsing.

In CASE 2, the FIR 6a cores are considered to have been produced by the interaction. 
In fact, \citet{Whitworth94} have demonstrated that 
the external shock compression could trigger the gravitational fragmentation 
of dense-gas clumps into cores. 
Then, we compare the separation among CORE 2, 3, and 4 with the Jeans length 
in order to investigate whether the fragmentation into the cores is caused by the gravitational instability. 
The mean projected separation ($\equiv$ $\Delta l$ ) among the FIR 6a cores, corresponding to the lower limit 
of the real separation, is $\sim$ 5 $\arcsec$ ($\sim$ 2.0 $\times$ 10$^{3}$ AU). 
The Jeans length ($\equiv$ $\lambda_J$) can be calculated with  
$\lambda_J=\sqrt{\frac{\pi C_{eff}^2}{Gn}}$ \citep{Naka07}, 
where $C_{eff}$, G, and $n$ are the effective sound speed, gravitational constant, 
and the average gas density of FIR 6a clump, respectively. 
We adopt a preshock density of $\sim$ 2.6 - 7.3 $\times$ 10$^{5}$ cm$^{-3}$ as an average gas density $n$
of FIR 6a clump before the interaction.
It is difficult, on the other hand, to observationally estimate $C_{eff}$, 
since there is no molecular emission, such as H$^{13}$CO$^{+}$ ($J$=1--0) and N$_2$H$^{+}$ ($J$=1--0), 
associated with FIR 6a clump \citep{Ike07, tate08}. 
Hence, we simply assume that $C_{eff}$ in FIR 6a clump is same as that in FIR 4 clump ($\sim$ 0.62 km s$^{-1}$).
With these values, the Jeans length $\lambda _J$ is estimated to be 5.0 - 8.4 $\times 10^3$ AU, 
which is comparable to the means separation among FIR 6a cores.
Then, it is possible that the interaction triggers the Jeans instability in FIR 6a clump,
which results in the fragmentation into FIR 6a cores. 
Furthermore, this interpretation is supported by the distribution of the dusty cores, which 
appears to delineate the outflow structure.

Our observational results cannot distinguish precisely between CASE 1 and CASE 2.
In either case, however,
FIR 6a clump contains three cores with a mass of 0.18 - 1.6 M$_{\odot}$ and a density of 
0.2 - 5.8 $\times 10^7$ cm$^{-3}$, 
and these cores may be potential formation sites of the next-generation of cluster members.

\end{bfseries}

\begin{figure}
   \begin{center}
      \FigureFile(160mm,100mm){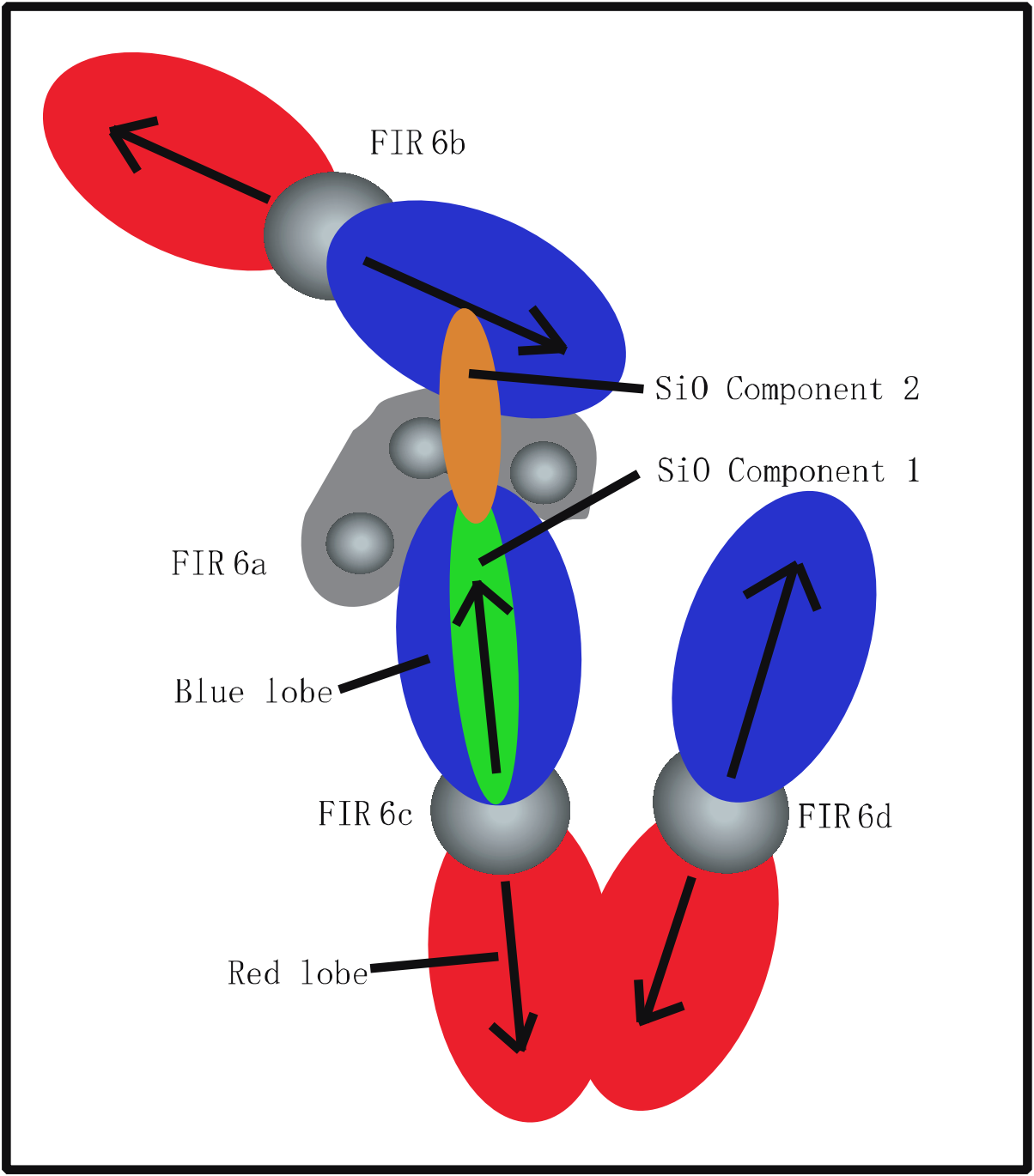}
   \end{center}
   \caption{Schematic picture in the FIR 6 region. 
}\label{overview}
\end{figure}

\begin{bfseries}

\section{SUMMARY}
 We have carried out high angular-resolution ($\sim$ 4$\arcsec$-7$\arcsec$) millimeter interferometric
 observations of the OMC-2 FIR 6 region
 with the NMA in the $^{12}$CO ($J$=1--0) and SiO ($v$=0, $J$=2--1) lines as well as the 3.3 mm continuum emission. 
The main results of our new millimeter observations are summarized as follows; 
\begin{enumerate}

\item{We detected dusty counterparts of FIR 6a-d in the 3.3 mm continuum emission. In particular
we have first resolved FIR 6a into three dusty cores and we totally detected six dusty cores. 
Typical size, mass, and the average gas density of these cores is estimated to be $\sim$ 1100-5900  AU, 
$\sim$ 0.18 - 5.5 M$_{\odot}$, and $\sim$ 10$^7$ cm$^{-3}$, respectively.}

\item {Our NMA observations in the $^{12}$CO ($J$=1--0) emission have confirmed the 
presence of molecular outflows
driven by FIR 6b and 6c, which were previously identified with 
Single-dish observations.
Furthermore, we have found a new outflow candidate whose driving 
source is FIR 6d.
Previous single-dish observations could not find this outflow, because 
this outflow
is hidden in the ambient cloud component.}

\item {We detected at least two well-collimated SiO ($v$=0, $J$=2--1) components aligned along
the axis of the blue lobe of the FIR 6c outflow, which probably traces the
well-collimated jet components ejected by FIR 6c. One of the SiO components,
Component 1, shows a higher blueshifted
velocity (-11.7 $\sim$ 11.0 km s$^{-1}$) than the other, Component 2 (3.4 $\sim$ 14.2 km s$^{-1}$),
while the tip of Component 2
locates more distant from the driving source. At the tip of Component 1,
the line width of the SiO emission shows abrupt increase and the
cores in FIR 6a clump form a shell-like feature,
suggesting the presence of the bowshock front. We consider that Component 1 is interacting with
FIR 6a clump while Component 2 is propagating to the different direction freely. 
Figure \ref{overview} shows a schematic picture in the FIR 6 region. }

\item{The estimated time scale of the fragmentation of FIR 6a clump into the three cores
($\sim$ 1.5 $\times$ 10$^{4}$ yr) is similar to the time scale 
of the interaction between the molecular outflow driven by FIR 6c and FIR 6a clump (1.2 - 1.9 $\times$ 10$^{4}$ yr),
and hence we cannot tell whether the fragmentation of FIR 6a clump into the cores occurs before or
after the interaction. In the former case, the interaction with the FIR 6c outflow is
unlikely to affect the subsequent evolution of the FIR 6a cores, since the momemtum flux of the
FIR 6c outflow is one order of magnitude smaller than the gravitational force in the FIR 6a cores.
In the latter case, it is possible that the interaction between the FIR 6c outflow and FIR 6a clump
triggered the fragmentation into the cores by the gravitational instability.
In either case, the FIR 6a cores might be potential sites of the next-generation cluster formation
in the FIR 6 region.
}

\end{enumerate}
\end{bfseries}

 We are grateful to the staff at the Nobeyama Radio Observatory (NRO) for both operating 
the NMA and helping us with the data reduction. Nobeyama Radio Observatory is a branch of 
the National Astronomical Observatory, National Institutes of Natural Sciences, Japan. 
We thank D. Johnstone for providing us the submillimeter continuum data taken with JCMT.
Moreover we acknowledge M. Yamada, N. Ikeda, Y. Kurono, and T. Tsukagoshi for their helpful comments. 
We also acknowledge the anonymous referee for providing helpful suggestions to 
improve the paper. Y. Shimajiri was financially supported by Global COE Program "the Physical Sciences Frontier", MEXT, Japan. 
S. Takahashi is supported by a postdoctoral fellowship of
the Institute of Astronomy and Astrophysics, Academia Sinica.
This work was supported by Grant-in-Aid for Scientific Research
A 18204017. S. Takakuwa acknowledges a grant from the National Science 
Council
of Taiwan (NSC 97-2112-M-001-003-MY2) in support of this work.

\clearpage

\clearpage

\clearpage

\clearpage

\clearpage

\clearpage

\clearpage

\clearpage
 \begin{table}
 \begin{center}
\caption{PARAMETERS FOR THE MOLECULAR LINE OBSERVATIONS}
 \label{line_obs}
 \begin{tabular}{lcc}
 \hline
 \multicolumn{1}{c}{Parameter}       & $^{12}$CO ($J$=1--0) & SiO ($v$=0, $J$=2--1) \\
 \hline
  Configuration\footnotemark[$*$]& \multicolumn {2}{c}{D} \\
  Baseline [k$\lambda$] & 3.9-31.4 & 3.3-27.0 \\
  Primary beam HPBW [arcsec] & 62 $\arcsec$ & 77 $\arcsec$ \\
  Synthesized Beam HPBW [arcsec] & 6${\farcs}$4 $\times$ 5${\farcs}$4 &  8${\farcs}$8 $\times$ 7${\farcs}$1 \\
  Velocity resolution [km s$^{-1}$] & 0.406 km s$^{-1}$ & 0.539 km s$^{-1}$ \\
  Gain calibrator\footnotemark[$\dagger$]& \multicolumn {2}{c}{0528+134} \\
  Bandpass calibrator\footnotemark[$\ddagger$] & \multicolumn{2}{c}{3C279}\\
  System temperature in DSB [K] & 200-300 K & 100-300 K \\
  Rms noise level [Jy beam$^{-1}$] & 0.18 Jy beam$^{-1}$ & 0.07 Jy beam$^{-1}$\\
 \hline
\multicolumn{3}{@{}l@{}}{\hbox to 0pt{\parbox{180mm}{\footnotesize
 \par\noindent
   \footnotemark[$*$] D is the most compact configuration.
  \par\noindent
  \footnotemark[$\dagger$] A gain calibrator, 0528+134, was observed every 20 minutes.
  \par\noindent
   \footnotemark[$\ddagger$] The response across the observed passband was determined from 40 minutes observations.
}\hss}}
  \end{tabular}
\end{center}
\end{table}

 \begin{table}
 \begin{center}
 \caption{PARAMETERS FOR THE CONTINUUM OBSERVATIONS}
 \label{cont_obs}
 \begin{tabular}{lcc}
 \hline
 \multicolumn{1}{c}{Parameter} & Figure \ref{dust}a & Figure \ref{dust}b   \\
 \hline
  Configuration & \multicolumn {2}{c}{C \& D}\\
  Baseline [k$\lambda$] & \multicolumn {2}{c}{2.5-53.1} \\
  Weighting & Natural & Uniform \\
  Beam size (HPBW) [arcsec] & 6${\farcs}$6 $\times$ 4${\farcs}$9  &  5${\farcs}$1 $\times$ 3${\farcs}$8 \\
  P.A. of the beam [$^\circ$] & -15.7  & -24.0 \\
  Gain calibrator $^*$ & \multicolumn{2}{c}{0528+134}  \\
  Bandpass calibrator $^\dagger$ & \multicolumn{2}{c}{3C84, 3C454.3} \\
  System temperature in DSB [K] & \multicolumn{2}{c}{100-300 K} \\
  Rms noise level [Jy beam$^{-1}$] & 9.7 $\times$ 10$^{-4}$ Jy beam$^{-1}$ & 1.2 $\times$ 10$^{-3}$ Jy beam$^{-1}$\\
 \hline
\multicolumn{3}{@{}l@{}}{\hbox to 0pt{\parbox{180mm}{\footnotesize
  \par\noindent
  \footnotemark[$*$]  A gain calibrator, 0528+134, was observed every 20 minutes.
  \par\noindent
   \footnotemark[$\dagger$] The response across the observed passband was determined from 40 minutes observations.
 }\hss}}
  \end{tabular}
\end{center}
\end{table}

 \begin{table}
 \begin{center}
 \caption{IDENTIFIED DUSTY CORES}
 \label{core}
 \begin{tabular}{lccccccc}
 \hline
 \multicolumn{1}{c}{} & &  & D$_{maj}$ $\times$ D$_{min}$$^*$ & P.A. & M$_{dust}$ $^\dagger$  & n  \\
  \multicolumn{1}{c}{source} &  $\alpha_{J2000}$ & $\delta_{J2000}$ & ($\times$10$^3$ AU) & (degree) & M$_{\odot}$ & (cm$^{-3}$)  \\
 \hline
  CORE 1 & 05$^h$ 35$^m$ 55$\fs$6 & -05$^\circ$ 12$\arcmin$ 3$\farcs$2  & 3.3 $\times$ 2.9  & 178.7 & 1.9 - 5.5  & 1.7 - 4.7 $\times$ 10$^{7}$  \\
  CORE 2 & 05 35 22.7 & -05 12 25.2 & 3.9 $\times$ 1.8 & 1.0 & 0.56 - 1.6 & 0.81 - 2.3 $\times$ 10$^{7}$ \\
  CORE 3 & 05 35 22.3 & -05 12 27.1 & 1.6 $\times$ 1.1  & 170.5 & 0.18 - 0.51 & 2.1 - 5.8 $\times$ 10$^{7}$ \\
  CORE 4 & 05 35 23.0 & -05 12 30.7 & 5.7 $\times$ 2.8 & 157.6 & 0.52 - 1.5  & 0.20 - 0.59 $\times$ 10$^{7}$ \\
  CORE 5 & 05 35 21.4 & -05 13 17.0 & 4.8 $\times$ 3.5  & 41.7 & 1.6 - 4.5 & 0.58 - 1.6 $\times$ 10$^{7}$ \\
  CORE 6 & 05 35 20.1 & -05 13 15.0 & 2.7 $\times$ 1.8  & 159.4 & 1.4 - 4.0 & 3.5 - 9.9 $\times$ 10$^{7}$ \\
 \hline
\multicolumn{7}{@{}l@{}}{\hbox to 0pt{\parbox{180mm}{\footnotesize
  \par\noindent
  \footnotemark[$*$] The FWHM size is estimated by the 2-dimensional Gaussian fitting to the uniformly-weighted image.
  \par\noindent
   \footnotemark[$\dagger$] Derived from the total flux density integrated within the 3$\sigma$ contour levels. See texts for details.
}\hss}}
  \end{tabular}
\end{center}
\end{table}

\end{document}